\documentclass[preprint2]{aastex}

\slugcomment{Submitted to {\sl The Astronomical Journal}}

\shorttitle{Radio continuum observations of NGC~4522}
\shortauthors{Vollmer et al.}

\begin{document}

\title{Radio continuum observations of the Virgo cluster spiral NGC~4522\\
The signature of ram pressure}

\author{B.~Vollmer\altaffilmark{1}}
\affil{
CDS, Observatoire astronomique de Strasbourg, UMR 7550, 11 rue de l'Universit\'e, 67000 Strasbourg, France}
\email{bvollmer@astro.u-strasbg.fr}

\author{R.~Beck}
\affil{Max-Planck-Institut f\"{u}r Radioastronomie, Auf dem H\"{u}gel 69, 53121 Bonn, Germany}
\email{rbeck@mpifr-bonn.mpg.de}

\author{J.D.P. Kenney}
\affil{Yale University Astronomy Department, P.O. Box 208101, New Haven, CT 06520-8101, USA}
\email{kenney@astro.yale.edu}

\and

\author{J.H. van Gorkom}
\affil{Department of Astronomy, Columbia University, 538 West 120th Street, New York, NY 10027, USA}
\email{jvangork@astro.columbia.edu}

\altaffiltext{1}{also at: Max-Planck-Institut f\"ur Radioastronomie, 
Auf dem H\"ugel 69, 53121 Bonn, Germany}

\begin{abstract}
Radio continuum observations at 20 and 6~cm of the highly inclined Virgo spiral
galaxy NGC~4522 are presented. Both, 20 and 6~cm total emission distributions are
asymmetric with an extended component to the west
where extraplanar atomic gas and H$\alpha$ emission are found. The 6~cm polarized
emission is located at the eastern edge of the galactic disk. Its peak is
located about 1~kpc to the east of the total emission peak.
We argue that this phenomena is a characteristic feature for cluster galaxies
which are experiencing significant pressure from the intracluster medium.
The degree of polarization decreases from the east to the west. The flattest
spectral index between 20 and 6~cm coincides with the peak of the 6~cm
polarized emission. These findings are consistent with a picture of a large
scale shock due to ram pressure located at the east of the galaxy
where cosmic rays are accelerated. We conclude that it is likely that the galaxy experiences 
active ram pressure.
\end{abstract}

\keywords{
ISM: clouds -- ISM: kinematics and dynamics -- Galaxies: individual: NGC~4522 -- Galaxies: evolution -- 
Galaxies: interactions -- Galaxies: ISM -- Galaxies: kinematics and dynamics
}

\section{Introduction}

The Virgo cluster spiral galaxy NGC~4522 is one of the few Virgo galaxies where
we can directly observe the effects of ram pressure due to the galaxy's rapid motion 
in the intra-cluster medium. 
Kenney \& Koopmann (1999) observed this galaxy in the optical and the H$\alpha$ line 
with the WIYN telescope. Whereas the highly inclined, old stellar disk appears symmetric, 
the distribution of ionized gas is highly asymmetric. The H$\alpha$ disk is
sharply truncated beyond 0.35\,$R_{25}$. Ten percent of the H$\alpha$ emission
arises from extraplanar H{\sc ii} regions that are exclusively located to
the west of the galactic disk. Kenney \& Koopmann (1999) argue that this
ionized gas distribution is reminiscent of a bow shock morphology, which suggests
that the gas is pushed to the west by ram pressure. 

Vollmer et al. (2000) obtained an H$\alpha$
velocity field with the Fabry-Perot Interferometer at the Observatoire de Haute Provence.
The galactic disk shows a symmetrically rising rotation curve. The velocities
of the extraplanar emission regions are not part of this regular rotation.
Their kinematic behaviour cannot be reproduced by rotation within a gravitational
potential of any known disk or halo model. Thus this gas is located out of the 
galactic plane and/or is accelerated/decelerated.

NGC~4522 has a large line-of-sight velocity with respect to the cluster mean 
($\sim$1300~km\,s$^{-1}$)
and is located at a distance of 3.3$^{\rm o}$ ($\sim$1~\footnote{We use a distance
of 17~Mpc to the Virgo cluster}Mpc) south from the Virgo cluster center (M87).
The intracluster medium (ICM) density at these distances might not be high enough to strip 
the gas at a
galactic radius of 0.35\,$R_{25}$. This is why Vollmer et al. (2000) argued that NGC~4522 has
been severely affected by ram pressure in the past and the extraplanar filaments represent 
gas that has been pushed to larger distances from the galaxy center and is now falling back 
onto the galactic disk.  

Recently, Kenney et al. (2003) observed NGC~4522 in the H{\sc i} 21~cm line with the VLA.
They found an atomic gas distribution that is similar to that of the ionized gas:
a truncated H{\sc i} disk with two high column density blobs in the north-west and the south-west.
The H{\sc i} velocity field of the extraplanar emission regions shows clear deviations from 
the overall rotation pattern. 

Even with the distributions and velocity fields of the atomic and ionized gas
it is still unclear if the galaxy is experiencing ongoing ram pressure stripping
or if the extraplanar gas is falling back to the galaxy after a past ram pressure
stripping event. On the one hand the galaxy's location and radial velocity 
excludes a simple radial orbit advocated by Vollmer et al. (2000), on the other hand
with the ongoing stripping scenario it is not clear if ram pressure is strong
enough at such a large distance from the cluster center where the ICM density is lower
than average.

Observations of the polarized radio continuum radiation can give important and
complementary information on the gas dynamics as has been shown for the barred
galaxy NGC~1097 (Beck et al. 1999). This radiation traces the ordered
large scale ($\sim$1~kpc) magnetic field in galactic ISM. This magnetic field 
is very sensitive to (i) compression and (ii) shear motion, which are both difficult
to detect in radial velocity fields. Thus, enhanced polarized radio continuum radiation
can be a sign of compression or strong shear motion in the gas. Both kinematic
features are present during a ram pressure stripping event (see e.g. Vollmer et al. 2000):
a galaxy on a highly eccentric orbit within the cluster experiences compression due to
ram pressure when it passes through the cluster core (compression phase);
after the core passage ram pressure ceases and the gas starts to fall back
onto the galactic disk, which gives rise to strong shear motions.

Observationally, there are only two Virgo cluster galaxies observed deeply in the 
polarized 6~cm radio continuum, because this is very time consuming (NGC~4254,
Soida et al. 1996, and NGC~4654, Chyzy et al. in prep.). Both galaxies show
an asymmetric distribution of polarized radio continuum emission.

In order to interpret the polarization data, numerical modeling is important, 
because different dynamical features, such as compression or shear, 
can produce polarized emission maxima.
Otmianowska \& Vollmer (2003) solved the induction equation using the
velocity fields of Vollmer et al. (2002) in order to calculate the evolution of the
large scale magnetic field during a stripping event. In a second step they 
calculated the polarized radio continuum emission and made maps of its evolution
that can be directly compared to observations. We will apply this method to
a dynamical model designed for NGC~4522 in a forthcoming paper.

In this article we present radio continuum observations at 6~cm and 20~cm obtained with
the VLA. The outline of the article is as follows. The observations are described
in Sec.~1. The results are discussed in Sect.~2. We compare our results to
radio observation of field galaxies (Sect.~4) and compare our data with existing
H{\sc i} and H$\alpha$ maps (Sect.~5). The discussion in Sect.~6 is followed by
the conclusions in Sect.~7.

\section{Observations}

NGC~4522 was observed on December, 26 and 31, 2001 with the
Very Large Array (VLA) of the National Radio Astronomy Observatory
(NRAO)\footnote{NRAO is a facility of National Science Foundation
operated under cooperative agreement by Associated Universities, Inc.}
D array configuration. The observation times were 9~h at 6~cm
and 1~h at 20~cm. The band passes were $2\times 50$~MHz at 6~cm and 20~cm.
We used 3C286 as flux calibrator and 1254+116
as phase calibrator, which was observed every 40~min.
The beam sizes of the final maps are $50'' \times 50''$ at 20~cm and
$15'' \times 15''$ at 6~cm. Solar activity caused interferences
at short baselines at 20~cm. Therefore, we had to remove a considerable amount of 
visibilities. This increased our final rms of the  20~cm map.
Maps where made for both wavelengths using the AIPS task IMAGR with ROBUST=0, which is 
in between pure uniform weighting (-5) and pure natural weighting (5).
We ended up with an rms of 150~$\mu$Jy/beam at 20~cm in total intensity and 
10~$\mu$Jy/beam at 6~cm in total and polarized emission. 

In addition, we use the H{\sc i} 21~cm line continuum of Kenney et al. (2003)
an effective bandpass of 68 channels e.g. 3.3~MHz.
NGC~4522 was observed during 6.7~h with the VLA CS array configuration, which lead to a
20~cm continuum rms of 160~$\mu$Jy/beam at a resolution of $\sim 20''$.

The total power flux of the low resolution 20~cm D array data is 24.3~mJy and
that of the H{\sc i} line continuum data is 25.2~mJy. Both values are
very close to the NVSS (Condon et al. 1998) flux of 22.3~mJy. The total power flux at 
6~cm is 7.6~mJy.

\section{Results}

\subsection{20~cm data \label{sec:20cm}}

Since the two sets of 20~cm continuum data have very different angular resolution, 
we present them separately.

\begin{figure}
	\resizebox{\hsize}{!}{\includegraphics{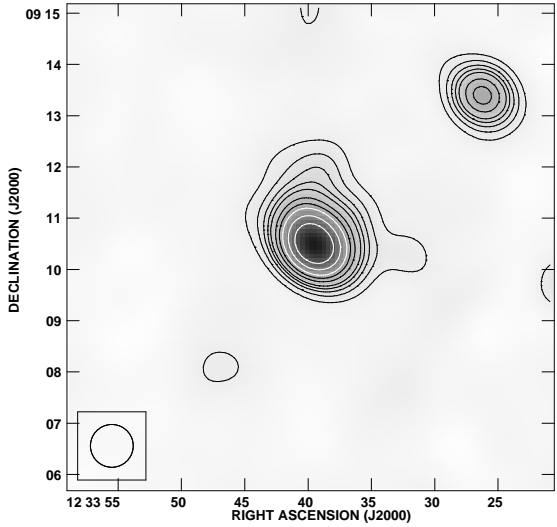}}
\figcaption[vollmer.fig1.ps]{20~cm D array total continuum emission of NGC~4522. The contour levels 
	are (1, 2, 3, 4, 5, 6, 7, 8, 9, 10)$\times$0.5~mJy.
	The beam ($50'' \times 50''$) is plotted in the lower left
	corner of the image.
\label{fig:n4522_20_TP}}
\end{figure}

The D array 20~cm total emission is shown in Fig.~\ref{fig:n4522_20_TP}. The disk
of NGC~4522 is clearly detected. The visible asymmetries in the north and
the south-west are due to point sources as can be seen in the
6~cm data (Sect.~\ref{sec:6cm}), which have a 3 times higher resolution.
Along the minor axis, the emission is slightly more extended to the west.

\begin{figure}
	\resizebox{\hsize}{!}{\includegraphics{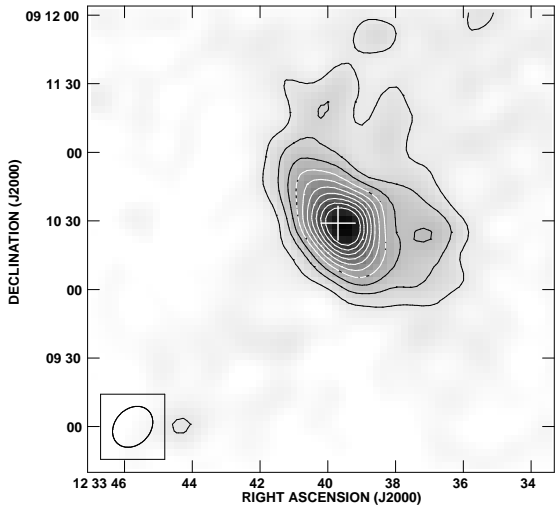}}
\figcaption[vollmer.fig2.ps]{20~cm H{\sc i} line continuum emission of NGC~4522. The contour levels 
	are (1, 2, 3, 4, 5, 6, 7, 8, 9, 10)$\times$0.48~mJy.
	The beam ($20'' \times 16''$) is plotted in the lower left
	corner of the image.
\label{fig:n4522_20_TP_CS}}
\end{figure}

The CS array H{\sc i} line continuum data are shown in Fig.~\ref{fig:n4522_20_TP_CS}.
The point source in the north of NGC~4522 is now clearly visible.
The continuum emission of NGC~4522 is asymmetric, i.e. it is more extended to the west.

However, we do not confirm the extended low surface brightness tail presented in a 
preliminary image by Kenney \& Koopmann (2001) to the west of the galaxy.

\subsection{6~cm data \label{sec:6cm}}

\begin{figure}
	\resizebox{\hsize}{!}{\includegraphics{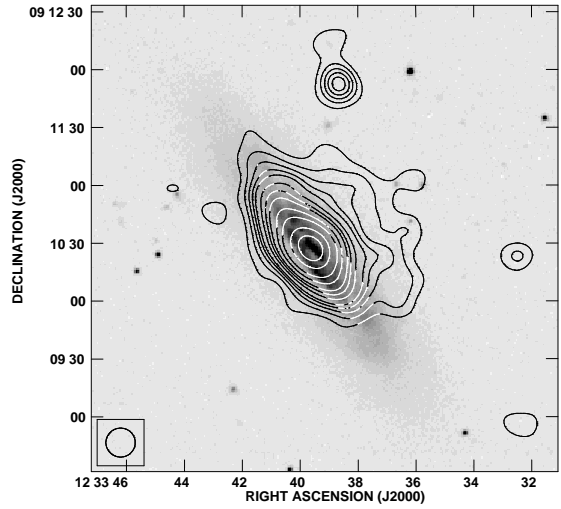}}
\figcaption[vollmer.fig3.ps]{6~cm total continuum emission of NGC~4522 as contours
        on an R band image (Kenney \& Koopmann 1999). The contour levels 
	are (1, 2, 3, 4, 5, 6, 7, 8, 9, 10, 20, 30)$\times$40~$\mu$Jy.
	The beam ($15'' \times 15''$) is plotted in the lower left
	corner of the image. 
\label{fig:n4522_opt_6TP}}
\end{figure}

The 6~cm total emission on an optical R band image (Kenney \& Koopmann 1999)  
is shown in Fig.~\ref{fig:n4522_opt_6TP}.
Two point sources, one $\sim 1.5'$ to the north and the other $\sim 1.7'$ to the west
of the galaxy center, are clearly visible 
(cf. Sect.~\ref{sec:20cm}). The emission from the disk is asymmetric,
i.e. it is more extended to the north-east. In addition, extended emission
in the west of the galactic disk is detected. 

\begin{figure}
	\resizebox{\hsize}{!}{\includegraphics{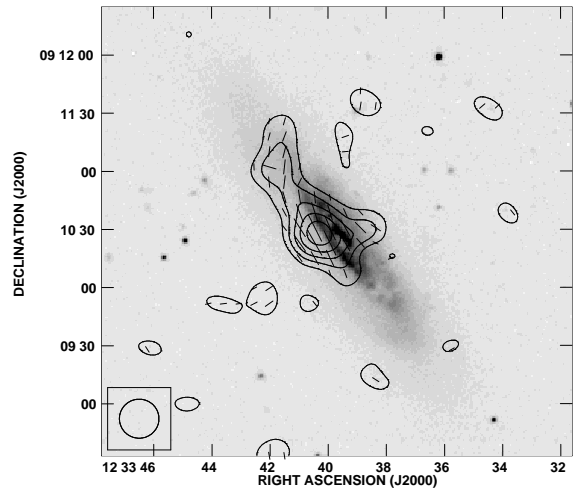}}
\figcaption[vollmer.fig4.ps]{6~cm polarized intensity as contours and the 
	vectors of the magnetic field uncorrected for
	Faraday rotation on an R band image (Kenney \& Koopmann 1999). 
	The size of the lines is proportional to the
	intensity of the polarized emission. The contour levels are
	(1, 2, 3, 4, 5)$\times$20~$\mu$Jy.
	The beam ($20'' \times 20''$) is plotted in the lower left
	corner of the image. 
\label{fig:n4522_opt_6PI_B}}
\end{figure}

The polarized 6~cm radio continuum emission together with the 
vectors of the magnetic field uncorrected for Faraday rotation on an R band image 
(Kenney \& Koopmann 1999) is shown in Fig.~\ref{fig:n4522_opt_6PI_B}.
In the absence of Faraday rotation, the magnetic field vectors are perpendicular
to the measured vectors of the electric field.
From experience with edge-on galaxies (Dumke et al. 2000) we expect
that Faraday rotation at 6~cm is smaller than 20$^{\rm o}$.
The polarized emission is clearly asymmetric. Its maximum is shifted 
12$''$ ($\sim$1~kpc) to the east and a few arcseconds to the north with respect 
to the peak of the total emission. It is elongated along the major axis of the galactic 
disk and is asymmetric. It extends mainly to the north-east.
There is also an elongation along the minor axis that extends
15$''$ to the north of the galactic center (the 6~cm total emission peak).
The magnetic field uncorrected for Faraday rotation is parallel to the 
galactic disk near the maximum of the polarized emission and bends northwards
to the north.

\begin{figure}
	\resizebox{\hsize}{!}{\includegraphics{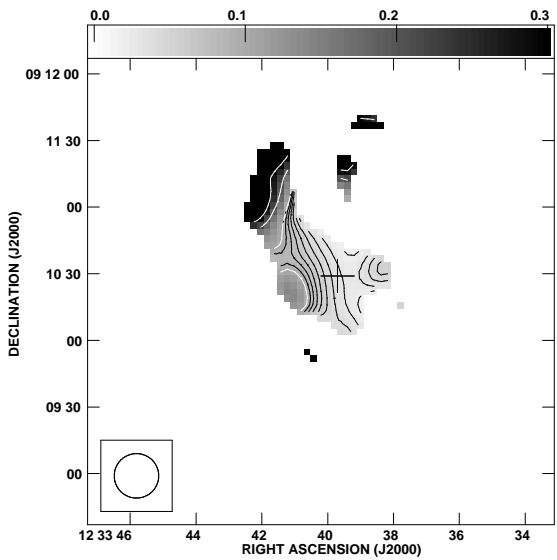}}
\figcaption[vollmer.fig5.ps]{The degree of polarization at 6~cm. The greyscale ranges
	from 0 to 0.3. The contour levels are (1, 2, 3, 4, 5, 6, 7, 8, 9, 10, 
	20, 30)$\times$0.01. The beam of the total and polarized emission
	($20'' \times 20''$) is shown in the lower left corner.
	The galaxy center is marked by a cross.
\label{fig:n4522_6poldeg}}
\end{figure}

Fig.~\ref{fig:n4522_6poldeg} shows the degree of polarization of the 
6~cm data. The 6~cm total emission map has been convolved to a beamsize
of 20$''$. The total emission has been clipped at 40~$\mu$Jy, the polarized
emission at 20~$\mu$Jy.
We observe a clear gradient in the west-east direction, with a rising degree of
polarization to the eastern edge of the galaxy. 
The degree of polarization along the minor axis decreases from the
east (13\%) to the west (3\%). The maximum of the degree of polarization (35\%)
is located at the northeastern edge of the galactic disk. 

\section{Comparison with field galaxies}

We find an asymmetric 6~cm polarized emission distribution
showing a ridge along the north-eastern edge of the galaxy.
Such polarized emission ridges are also observed in two other Virgo
spiral galaxies: NGC~4254 (Soida et al. 1996) and NGC~4654 (Chyzy et al., in prep.).
Are these asymmetric polarized emission ridges at the edges of the galaxies
characteristic for cluster galaxies or can we also observe them in field
galaxies? 

First, we compare the polarized emission distribution of NGC~4522 to that of
highly inclined field galaxies. 
Polarized radio continuum emission data at 20~cm  exist for NGC~4631 
(Hummel et al. 1988), NGC~891 (Hummel et al. 1991), NGC~4565 (Sukumar \& Allen 1991),
and NGC~5907 (Dumke et al. 2000). Whereas NGC~4631, NGC~891, and NGC~4565
show fairly symmetric polarized emission distributions, NGC~5907 
has a polarized ridge in the south-west. However, one has to take into account
that these data suffer from severe Faraday depolarization, because
of the long line of sight through the whole galactic disk. Thus one
mainly sees the part of the galaxy which is closer to the observer.
This effect already causes an asymmetry along the minor axis. Spiral
arms can then cause asymmetries along the major axis. Therefore,
it is necessary to observe these galaxies at shorter wavelengths, where
Faraday depolarization is small. Sukumar \& Allen (1991) observed the
nearby edge-on galaxy NGC~891 at 6~cm with the VLA. They found a
fairly symmetric double peaked distribution with a more prominent 
north-eastern peak. The degree of polarization rises up to 32\% to the
edges of the galaxy. Dumke et al. (2000) observed NGC~5907 at 6~cm
in polarization. They found that, despite the asymmetric distribution of the
polarized emission at 20~cm, the 6~cm polarized emission distribution is
almost symmetric. 
Thus, we conclude that highly inclined field galaxies can have heavily asymmetric
polarized emission distributions at 20~cm, but not at 6~cm.

Second, we compare the polarized emission distribution of NGC~4522 to 
that of field galaxies with smaller inclination angles (more face-on).
A compilation of imaging polarization measurements of 17 nearby galaxies 
at 6~cm with the VLA, ATCA, and the Effelsberg 100m telescope is available on
the website of the Max-Planck-Institut f\"{u}r Radioastronomie 
Bonn\footnote{http://www.mpifr-bonn.mpg.de/staff/wsherwood/mag-fields.html}.
None of the non-interacting, isolated galaxies (NGC~2276 has a companion,
M~101 is a member of a group) show prominent maxima or ridges 
of 6~cm polarized radio continuum emission at the outer edge of the galactic disk as
it is observed for NGC~4254, NGC~4654, and NGC~4522.
We conclude that these maxima are characteristic for cluster
spiral galaxies.

\section{Comparison with other wavelengths}

\subsection{Total emission}

\begin{figure}
	\resizebox{\hsize}{!}{\includegraphics{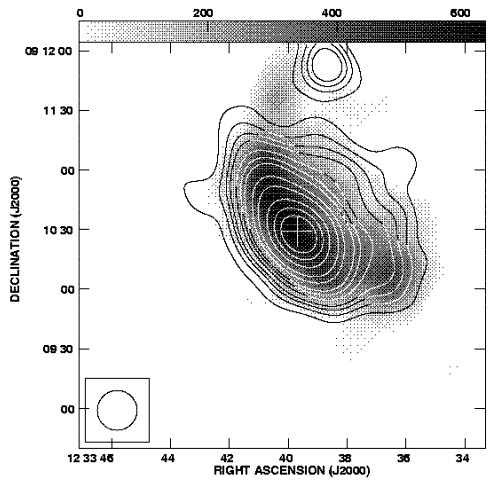}}
\figcaption[vollmer.fig7.ps]{Greyscale: H{\sc i} gas distribution (Kenney et al. 2003).
	Contours: 6~cm total emission. Both images are convolved to a 
	resolution of 20$''$. The galaxy center is marked by a cross.
\label{fig:n4522_TP_HI}}
\end{figure}

In order to investigate the relationship between the radio continuum emission distribution
on the one hand and the distributions of massive star formation and atomic gas on
the other hand, we compare our data with the H$\alpha$ data of Kenney \& Koopmann (1999) 
(Fig.~\ref{fig:n4522_Ha_TP_conv}) and the H{\sc i} map of Kenney et al. (2003)
(Fig.~\ref{fig:n4522_TP_HI}). All images have been convolved to the
same resolution.  
The total 6~cm disk emission is very similar to that of the H$\alpha$ emission,
except that its northern part deviates much less from the major axis
than the H$\alpha$ emission. Moreover, the radio emission is more extended
to the east along the minor axis. The radio data also show
extraplanar emission to the west. Whereas both, H{\sc i} and H$\alpha$
emission, are weak along the minor axis relative to their strength in the
south-western extraplanar peak, the radio continuum emission is equally strong 
at the minor axis and at the location of the south-western extraplanar H{\sc i} 
and H$\alpha$ peak.

The total 6 cm emission coincides with the atomic gas distribution
in the southern part of the galaxy and in the large western extraplanar peak.
In the north the radio emission deviates less from the major axis
than the atomic gas distribution. However, the H$\alpha$ emission shows the
same deviation from the major axis as the H{\sc i} gas distribution,
which extents much further out.
Thus, along the minor axis and in the north of NGC~4522 the 6~cm total 
continuum emission deviates from both, the H$\alpha$ and H{\sc i} emission distribution.

\subsection{Polarized emission}

\begin{figure}
	\resizebox{\hsize}{!}{\includegraphics{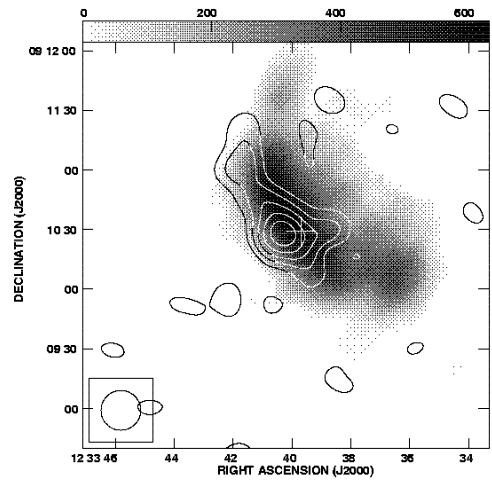}}
\figcaption[vollmer.fig8.ps]{Greyscale: H{\sc i} gas distribution (Kenney et al. 2003)
	with a resolution of $\sim$20$''$. Contours: 6~cm polarized
	radio continuum emission. The galaxy center is marked by a cross.
\label{fig:n4522_PI_HI}}
\end{figure}

An overlay of the 6~cm polarized radio continuum emission on the H{\sc i}
gas distribution is shown in Fig.~\ref{fig:n4522_PI_HI}.
The polarized emission ridge traces the outer north-eastern edge of the
gas distribution.

\begin{figure}
	\resizebox{\hsize}{!}{\includegraphics{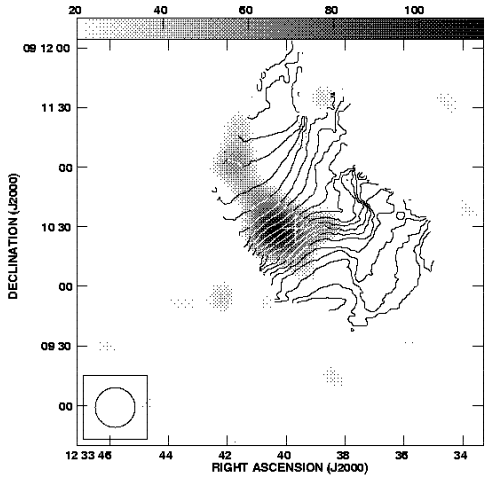}}
\figcaption[vollmer.fig9.ps]{Greyscale: 6~cm polarized radio continuum emission.
	Contours: H{\sc i} velocity field (Kenney et al. 2003).
	The contour levels are from 2150~km\,s$^{-1}$ to 2440~km\,s$^{-1}$
	in steps of 10~km\,s$^{-1}$. The beam of the H{\sc i} data can be seen
	in the lower left corner. The galaxy center is marked by a cross.
\label{fig:n4522_PI_HIvel}}
\end{figure}

The strong polarized emission can be due to an enhanced regular magnetic field or
an increased density of cosmic ray electrons. Since the total radio emission
is not enhanced in this region, the latter possibility can be excluded.
In order to investigate the link between the enhancement of the regular magnetic field
and the galaxy kinematics, the 6~cm polarized radio continuum emission
is plotted on the H{\sc i} velocity field in Fig.~\ref{fig:n4522_PI_HIvel}.
The velocity field in the region of the maximum of the polarized emission
is very regular. 
The most interesting feature is located to the west of the galactic center.
There, the polarized emission lies at the edge of a region of strong kinematic
perturbation. This region not only shows bent isovelocity contours,
but also an unusually large linewidth of $\sim$100~km\,s$^{-1}$ (Kenney et al. 2003). 
Shear velocities in the perturbed regions of the velocity field might be responsible 
for the western extension of the polarized radio emission.

\section{Discussion}

\begin{figure}
	\resizebox{\hsize}{!}{\includegraphics{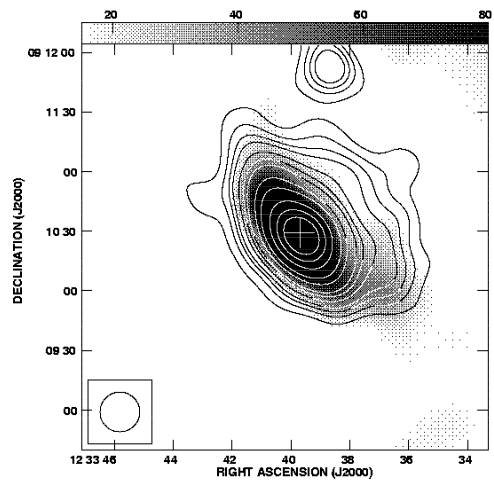}}
\figcaption[vollmer.fig6.ps]{Greyscale: H$\alpha$ emission distribution (Kenney \& Koopmann 1999). 
	Contours: 6~cm total emission. Both images are convolved to a 
	resolution of 20$''$. The galaxy center is marked by a cross. 
\label{fig:n4522_Ha_TP_conv}}
\end{figure}

\begin{figure}
	\resizebox{\hsize}{!}{\includegraphics{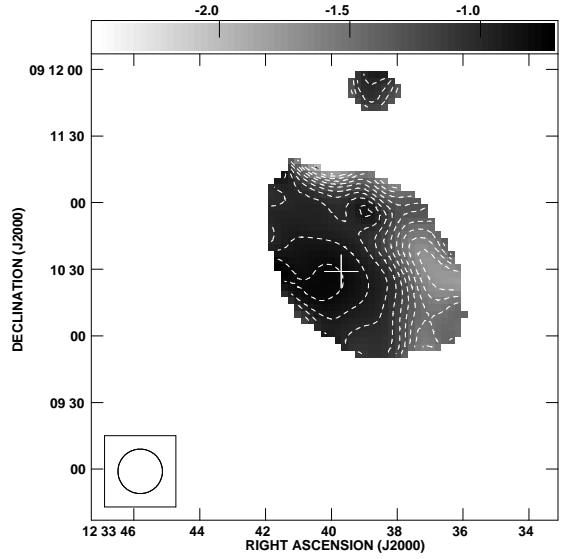}}
\figcaption[vollmer.fig10.ps]{Spectral index map between 20~cm (Kenney et al. 2003) and our 6~cm data.
	The contours range from -2.3 to -0.7 in steps of 0.1.
	The galaxy center is marked by a cross. 
\label{fig:n4522_specindex}}
\end{figure}

Both 6~cm and  20~cm radio continuum maps show a symmetric disk together with a region  
of extended emission to the west.
The 6~cm radio continuum emission distribution follows the H{\sc i} gas distribution
(Fig.~\ref{fig:n4522_TP_HI}) more closely than the distribution of massive star formation 
(H$\alpha$, Fig.~\ref{fig:n4522_Ha_TP_conv}). In the south-western extraplanar gas
H$\alpha$ emission is concentrated in the southern part and its maximum does not
coincide with the H{\sc i} emission maximum. In the region between the two
extraplanar gas maxima only a small amount of H$\alpha$ emission is
observed, but radio continuum emission is clearly detected. 
(Fig.~\ref{fig:n4522_Ha_TP_conv}). This can be for two reasons: (i) 
extinction affecting the H$\alpha$ data or (ii) the radio data traces
electrons that diffuse into the the region between the two extraplanar gas maxima.
In order to discriminate between the two possibilities we show the spectral index 
map between 20~cm (Fig.~\ref{fig:n4522_20_TP_CS}) and our 6~cm data in Fig.~\ref{fig:n4522_specindex}.
Both maps were convolved to a resolution of $20'' \times 20''$.
In the discussed region the spectral index is steeper than -1. This excludes a
hidden star formation region which is not visible in the H$\alpha$ map, because
such a region would show a shallower spectrum due to an increased thermal electron
population. The cosmic ray gas that escapes from the galactic disk expand leading
to a decreasing cosmic ray particle density away from the disk. Since the synchrotron
emission depends on the cosmic ray particle density, the radio emission from
magnetized gas is stronger if it is located close to the disk. We thus conclude
that the total 6~cm radio emission in the region between the northern
and southern extraplanar H{\sc i} maxima is due to diffusion of accelerated
electrons suggesting that this atomic gas is located close to the galactic disk.

The northern extraplanar atomic gas is not seen in the 6~cm total radio continuum 
emission. We suggest that this is, because it is located farther away from the
galaxy plane than the region between the two extraplanar gas maxima.

The spectral index map (Fig.~\ref{fig:n4522_specindex})
clearly shows a steepening of the spectral index to the west
from -1 in the disk to values smaller than -2 in the south-western extraplanar gas.
This is consistent with diffusion and aging of the electrons producing the synchrotron emission
out there. Since the radio spectrum of the galactic disk emission 
is already relatively steep, the fraction of thermal electrons cannot exceed $\sim10-20$\%.
Thus, away from the disk it is unlikely that the steepening of the spectral index is due to a
variation of the fraction of thermal emission.

\begin{figure}
	\resizebox{\hsize}{!}{\includegraphics{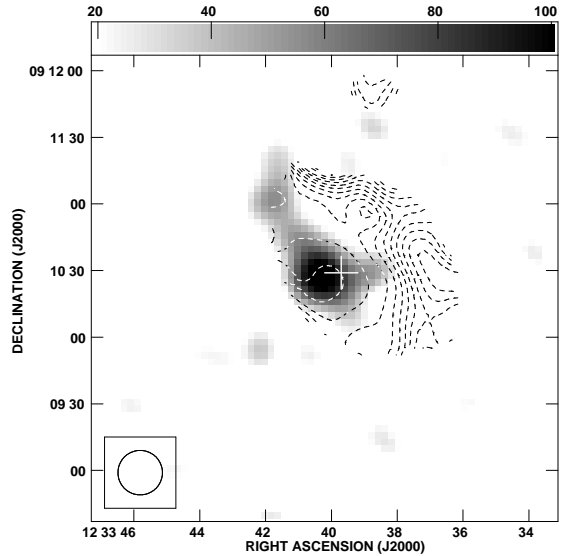}}
\figcaption[vollmer.fig11.ps]{Spectral index map between 20~cm (Kenney et al. 2003) and our 6~cm data
	as contours on the polarized radio continuum emission.
	The galaxy center is marked by a cross. 
\label{fig:n4522_spin_PI}}
\end{figure}

An even more remarkable effect is that the flattest part of the index distribution
does not coincide with the H$\alpha$/total radio emission peak, i.e. the galactic center,
but with the peak of the polarized radio emission as it can be seen in Fig.~\ref{fig:n4522_spin_PI}.
To our knowledge such a shift has never been observed before. 
In general the flattest part of the spectral index distribution coincides
with that of the radio emission peak (see, e.g., Duric et al. 1998). A possible interpretation 
would be that the electrons are accelerated to relativistic velocities 
in the large scale (kpc) shock that produces the
eastern polarized radio emission ridge (Duric 1986). The radio emission of such an
electron population has a spectral index of -0.5 (Bell 1978). 
The observed spectral index of -0.7 would then be due to subsequent energy loss 
via synchrotron emission.

The enhanced polarization proves that the radio continuum ridge is
due to a shock and not to enhanced star formation. Star formation
would lead to enhanced turbulent motions that would destroy the ordered magnetic field.
In addition, the energy density of the ordered magnetic field is at least an order of
magnitude smaller than that of the turbulent magnetic field, because the total
power emission is not significantly enhanced in this region.

Polarized radio continuum emission is a powerful tool to 
detect interactions of spiral galaxies with the cluster ICM. 
In the case of NGC 4522 the shock may be due to the compression
of the ISM due to ram pressure. An alternative scenario 
where the shock would be caused by stripped gas falling back 
onto the galactic disk from the north (Otmianowska-Mazur \& Vollmer 2003) is no
longer viable. The HI gas distribution and kinematics  (Kenney et al 2003) and our
radio contiuum observations rule out that the gas is currently falling back.
If the shock is caused by ongoing ram pressure the galaxy
must be moving eastward in the plane of the sky. If one assumes trailing spiral arms,
the eastern edge of NGC~4522 is located closer to the observer. The galaxy then rotates
counter-clockwise. The blueshift of the northern extraplanar gas maximum (Kenney et al. 2003)
and our conclusion (that it is located far away from the galactic plane)
are consistent with this picture, because this part of the ISM is
accelerated by ram pressure.

\subsection{Future work}

In order to corroborate our scenario of active ram pressure stripping, VLA observations at 
3.6~cm will be needed. With these data, (i) a more reliable spectral index map can be made, (ii) 
thermal and non-thermal emission can be separated, and
(iii) a rotation measure map can be obtained. In the case of active ram pressure
stripping one expects a constant or increasing rotation measure along the minor axis from the
east to the west, because the stripped gas is pushed to the west.
Since the eastern edge of the galaxy is closer to the observer, the stripped, partially ionized 
gas is located between galaxy's disk and the observer and thus acts as a Faraday screen.
The absence of Faraday rotation would speak against the active stripping scenario.

Polarized radio continuum emission is a powerful tool to detect interactions
of spiral galaxies with their cluster environment. A systematic polarization
survey of cluster spiral galaxies can give us unique
informations about ISM--ICM interactions. When an asymmetric polarized radio 
emission distribution is detected, the direction of the galaxy's tangential velocity 
can be determined and its absolute value estimated. The combination of the
H{\sc i} and radio continuum data allows us to estimate the galaxy's 3D distance
from the cluster center. Polarized radio continuum data thus give important and
complementary information to H{\sc i} and H$\alpha$ observations.

\section{Conclusions}

We present VLA D array observations at 6 and 20~cm of the highly inclined Virgo
spiral galaxy NGC~4522. The results are
\begin{enumerate}
\item
the 20~cm and 6~cm total emission distributions are asymmetric; they are more extended to
the west, where extraplanar H$\alpha$ and H{\sc i} emission is found,
\item
the 6~cm polarized emission is located at the eastern edge of the galactic disk;
its peak is displaced to the east of the total emission peak,
\item
the degree of polarization increases from the galaxy center towards the eastern 
peak and towards the northeastern edge of the disk,
\item
the spectral index between 20 and 6~cm decreases from west to east and its maximum
coincides with the peak of the 6~cm polarized emission,
\item
polarized radio continuum emission is a powerful tool to detect interactions
of spiral galaxies with their cluster environment.
\end{enumerate}
Points 3 and 4 are consistent with a picture of a large
scale shock due to ram pressure located at the east of the galaxy
possibly associated with particle acceleration.
We argue that the asymmetry of the 6~cm polarized emission is characteristic 
for cluster galaxies and is a sign of interaction with its cluster environment. 
We conclude that it is probable that the galaxy experiences active
ram pressure, but it is not clear whether we are observing it before, during, or after peak ICM 
pressure.

\begin{acknowledgements}
This work was supported in part by NSF grant AST-00-98294 to Columbia University
and NSF grant AST-00-71251 to Yale University.
\end{acknowledgements}

\clearpage

\begin{thebibliography}{}

\bibitem{q1} Beck R., Ehle M., Shukurov A., \& Sokoloff D. 1999, Nature, 397, 324 

\bibitem{q2} Bell A.R. 1978, MNRAS, 182, 147

\bibitem{q3} Condon J.J., Cotton W.D., Greisen E.W., et al. 1998, AJ, 115, 1693

\bibitem{q4} Dumke M., Krause M., \& Wielebinski R. 2000, A\&A, 355, 512

\bibitem{q5} Duric N. 1986, A\&A, 304, 111 

\bibitem{q6} Duric N., Irwin J., \& Bloemen H 1998, A\&A, 331, 428

\bibitem{q7} Hummel E., Lesch H., Wielebinski R., \& Schlickeiser R. 1988, A\&A, 197, L29

\bibitem{q9} Kenney J.P.D. \& Koopmann R.A. 1999, AJ, 117, 181

\bibitem{q10} Kenney J.P.D. \& Koopmann R.A. 2001, in: Gas \& Galaxy Evolution, ASP Conference Series, ed: J.E. Hibbard, M.P. Rupen, and J.H. van Gorkom, Vol. 240, p.577

\bibitem{q11} Kenney J.P.D., van Gorkom J., \& Vollmer B. 2003, ApJ, submitted

\bibitem{q12} Kotanyi C., van Gorkom J.H., \& Ekers R.D. 1983, ApJ, 273, L7

\bibitem{q13} Otmianowska-Mazur K. \& Vollmer B. 2003, A\&A, 402, 879

\bibitem{q14} Soida M., Urbanik M. \& Beck R. 1996, A\&A, 312, 409

\bibitem{q15} Sukumar S. \& Allen R.J. 1991, ApJ, 382, 100

\bibitem{q16} Vollmer B., Marcelin M., Amram P., Balkowski C., Cayatte V., \& Garrido O. 
2001, A\&A 364, 532

\bibitem{q17} Vollmer B., Cayatte V., Balkowski C., \& Duschl W.J. 2001, ApJ, 561, 708

\end{thebibliography}
\end{document}